\documentclass[11pt,a4paper]{article}
\usepackage{amsmath,amssymb}
\usepackage{epsfig,graphicx}

\begin{document}

\title{\bf Dark halos built of scalar gravitons: numerical study}
\author{Yu.~F.~Pirogov$^{a,\, b}$\footnote{{\bf e-mail}: yury.pirogov@ihep.ru},
I.~Yu.~Polev$^{b}$\footnote{{\bf e-mail}: igor.polev@gmail.com}
\\
$^a$ \small{\em Theory Division, Institute for High Energy Physics} \\
\small{\em 
 Protvino, Moscow Region, Russia}
\\
$^b$ \small{\em  Faculty of General and Applied Physics, Moscow Institute of
Physics and Technology} \\
\small{\em 
Dolgoprudny, Moscow Region, Russia}
}
\date{}
\maketitle

\begin{abstract}
\noindent
In  a previous article due to one of the present authors (YFP),  an extension to
General Relativity,  violating  general
covariance to the residual unimodular one, was proposed.  As a manifestation of
such   a violation,  there appears  the (massive)
scalar  graviton  in addition to the massless tensor one. The former was
proposed as  a candidate on the dark matter in the Universe.  
In a subsequent  article (Yu. F. Pirogov,  MPLA 24, 3239, 2009;
arXiv:0909.3311 [gr-qc]),  an application
of the extension was developed. Particularly, a regular solution to the static
spherically symmetric equations   in  empty space was studied by means of
analytical methods.  This  solution was proposed as a prototype model for
the galaxy soft-core  dark halos, with the coherent
scalar-graviton field as dark matter.  The  present report is a supplement to
the aforementioned article. The statements of  the latter are  verified and 
visualized by means of   numerical analysis and symbolic calculations. The
nice validity of  analytical results is found. 

\end{abstract}

\section{Introduction}

In Ref.~\cite{Pir0}, an extension to General Relativity  was proposed.  The
extension possesses the residual unimodular covariance, and  in line with the
massless tensor  graviton describes the (massive) scalar one.   The latter was
proposed as a candidate on  dark matter in the Universe. 
The theory was further developed in a series of  subsequent articles.
In particular, in~\cite{Pir} a regular solution to the static spherically
symmetric equations   of  extended gravity in  empty space 
was qualitatively studied by means of  analytical methods. This
solution  was proposed as a prototype model for the galaxy soft-core dark 
halos, with the coherent scalar-graviton field as dark matter.  
For details, we refer the reader to~\cite{Pir0,Pir}.
The  present report is a supplement to Ref.~\cite{Pir}. The statements of  the
latter are  verified and visualized by means of   numerical analysis  and
symbolic calculations.
The nice consistency of  the qualitative analytical study  is found.
As a by-product, it is found highly plausible that the power series for the
regular solution has just a finite radius of convergence.

\section{Extended gravity equations}

Here, we shortly remind  the results of~\cite{Pir}  concerning the static
spherically symmetric equations of the unimodular extended gravity in  empty
space. The line element in such a case  looks in the polar coordinates
$(\tau, r, \theta,\varphi)$ most generally like:
\begin{equation}\label{ds2}
ds^2  = ad\tau^2  - bdr^2  - cr^2 (d\theta ^2  + \sin ^2 \theta d\varphi ^2 ),
\end{equation}
where $a$, $b$ and $ c$ are some metric potentials. 
In the static limit, all the
variables depend only on the radius $r$, which is still defined ambiguously. 
To eliminate such an  ambiguity impose the coordinate condition 
 $ab=1$. 
Defining now the new variables 
\begin{equation}
 A \equiv a = 1/b,\  \ C \equiv r^2 c
\end{equation}
and supplementing them by   the (dimensionless) scalar-graviton  field $X$
we get the system of the nonlinear differential equations as follows:
\begin{eqnarray}
\frac{d}{{d r}}\left( {AC\frac{d}{{d r}}X} \right) &=& \frac{6}{R_h^2} C\exp
\left( { - X} \right),\nonumber\\
\frac{d}{{d r}}\left( {C\frac{d}{{d r}}A} \right) &=&
\frac{6 \varepsilon^2_h}{R_h^2}  C\exp \left( { - X} \right),\nonumber\\
\frac{d}{{d r}}\left( {C\frac{d}{{d r}}C} \right) - \frac{3}{2}\left(
{\frac{d}{{d r}}C} \right)^2 & =&  - \frac{{\varepsilon^2 _h }}{{2 }}\left(
{C\frac{d}{{d r}}X} \right)^2, \nonumber\\
\frac{d}{{d r}}\left( {C\frac{d}{{d r}}A} \right) - \frac{d}{{d r}}\left(
{A\frac{d}{{d r}}C} \right) + 2& =& 0.
\end{eqnarray}
Here,  a dimensionless parameter  $\varepsilon ^2 _h  = 2{k^2 _h }/{k_g ^2 }$
refers to  the  Lagrangian of extended gravity, 
with $\kappa_g$ being the mass scale for the ordinary gravity (of the order of
Planck mass) and  $\kappa_h$ being an additional mass scale
characteristic of the unimodular extended gravity.   On the contrary, the
parameter $R_h> 0$, with  the dimension of length,  arises  as a free
integration constant.  Physically, it characterizes  an internal length scale
for  a vacuum solution.  
 
The first equation above reflects
continuity condition in  empty space, with the other equations being a
combination of the continuity condition  and  three gravity equations. Out
of four equations, only three  are  independent and  can  be chosen at will as
a primary system. The remaining equation should, generally, 
serve as a consistency condition.

At a finite $R_h$,   it is appropriate to choose the scaled distance $\xi
=r/R_h$ as an independent variable. Introducing then $t=\xi^2$  and 
redefining $C$ as $C\equiv t c$ we get equivalently:
\begin{eqnarray}\label{MGE}
2t\frac{d}{{dt}}\left( {AC\frac{d}{{dt}}X} \right) + AC\frac{d}{{dt}}X& =
&3C\exp \left( { - X} \right), \nonumber\\
2t\frac{d}{{dt}}\left( {C\frac{d}{{dt}}A} \right) + C\frac{d}{{dt}}A &=&
3\varepsilon ^2 _h C\exp \left( { - X} \right), \nonumber\\
2 C\frac{{d^2 }}{{dt^2 }}C + \frac{1}{t}C\frac{d}{{dt}}C - \left(
{\frac{d}{{dt}}C} \right)^2 & =&  - \varepsilon ^2 _h \left( {C\frac{d}{{dt}}X}
\right)^2, \nonumber\\ 
2t\left( {C\frac{{d^2 }}{{dt^2 }}A - A\frac{{d^2 }}{{dt^2 }}C} \right) + \left(
{C\frac{d}{{dt}}A - A\frac{d}{{dt}}C} \right) + 1& = &0.
\end{eqnarray}
This system is the main concern of  the present  investigation.

\section{Analytical study}

Here, we recapitulate some results of~\cite{Pir} concerning the solution
regular  at  $t=0$. Looking for such a solution   as a
 power series of  $t$ one gets:
\begin{eqnarray}\label{analyt}
\tilde X &=& t  - \frac{1}{2}\left( {\frac{3}{5} + \varepsilon ^2 _h }
\right)t^2 + \left( {\frac{1}{{35}}\left( {4 + \frac{{41}}{3}\varepsilon ^2 _h }
\right) +
\frac{1}{3}\varepsilon ^4 _h } \right)t^3, \nonumber\\
\tilde a &=& 1 + \varepsilon ^2 _h \left( {t  - \frac{3}{{10}}t ^2  +
\frac{1}{{35}}\left( {4 
+ \frac{{19}}{6}\varepsilon ^2 _h } \right)t ^3 } \right),\nonumber\\
\tilde c &=& 1 + \varepsilon ^2 _h \left( { - \frac{1}{{10}}t ^2  +
\frac{2}{7}\left(
{\frac{1}{5} 
+ \frac{1}{3}\varepsilon ^2 _h } \right)t ^3 } \right).
\end{eqnarray}
This representation is, generally,  valid  at the arbitrary $\varepsilon_h\leq
1$ and may  formally be continued up to the  arbitrary powers of $t$. But being
restricted   just to  a region of $t$ (see later)  it does not give 
apprehension of  solution as a whole. 

To proceed,  we restrict ourselves  to the case 
$\varepsilon_h\ll1$ which is preferred from  astronomical observations.
Under this assumption, decompose formally an arbitrary  solution as  a power
series of $\varepsilon _h^2$:
\begin{equation}
X(t) = \sum\limits_{n = 0}^\infty  {\varepsilon _h ^{2n} } X_n (t),\ 
\ a(t) = \sum\limits_{n = 0}^\infty  {\varepsilon _h ^{2n} } a_n (t),\
\ c(t) = \sum\limits_{n = 0}^\infty  {\varepsilon _h ^{2n} } c_n( t),
\end{equation}
with the  conditions  $a_0  = c_0 = 1$.
Substituting this decomposition into (\ref{MGE}) we simplify the latter in the
leading $\varepsilon_h$-order  as follows:
\begin{eqnarray}\label{MGE0}
\frac{d}{{dt}}X_0  + \frac{2}{3}t\frac{{d^2 }}{{dt^2 }}X_0 & =& \exp \left(  -
X_0  \right),\nonumber\\
\frac{d}{{dt}}a_1  
+ \frac{2}{3}t\frac{{d^2 }}{{dt^2 }}a_1 & =&    \exp \left( { -
X_0 } \right),\nonumber\\
\frac{3}{t}\frac{d}{{dt}}c_1  + 2\frac{{d^2 }}{{dt^2 }}c_1 & =&  - \left(
{\frac{d}{{dt}}X_0 } \right)^2,\nonumber\\
 a_1  - t\frac{d}{{dt}}a_1  - 2t^2 \frac{{d^2 }}{{dt^2 }}a_1  
&=&  c_1  + 5t\frac{d}{{dt}}c_1  + 2t^2 \frac{{d^2 }}{{dt^2 }}c_1.
\end{eqnarray}
Clearly,  $a_1=X_0$ under the proper boundary conditions.
Thus,  solving the system of the
coupled differential equations~(\ref{MGE}) reduces  to solving the ordinary
differential equation for $X_0$ and $c_1$, with the last equation of
system  serving as a constraint.

To this end, introducing the  new  variables 
\begin{equation}
Z = X_0  - \sigma,\ \ \sigma  = \ln \left( {3t} \right)
\end{equation}
transform   the equation for $X_0$ to the autonomous  (not containing
explicitly the independent variable) form 
\begin{equation}
\frac{{d^2 }}{{d\sigma ^2 }}Z + \frac{1}{2}\frac{d}{{d\sigma }}Z =
\frac{1}{2}\left( {\exp ( { - Z} ) - 1} \right).
\end{equation}
Putting  then $\dot Z\equiv d Z/d \sigma$ reduce the second-order equation for
$Z$ to   system of  two first-order equations for $Z$ and $\dot Z$ as
follows: 
\begin{eqnarray}\label{1order}
\frac{d}{{d\sigma }}Z & = & \dot Z \nonumber\\
\frac{d}{{d\sigma }}\dot Z & = &  - \frac{1}{2}\dot Z + \frac{1}{2}\left( {\exp
\left( { - Z} \right) - 1} \right).
\end{eqnarray}

Such a system is known to be fully characterized by its  phase plane $(Z,\dot
Z)$. Of the particular  importance are the exceptional points   defined by
$dZ/\sigma=d \dot Z/d\sigma =0$. In the case at hand, there is 
just one point of this kind, $ Z=\dot Z=0$,  and it belongs to 
the stable focus type.  All the trajectories   $(Z(\sigma), \dot
Z(\sigma))$ winds around this point approaching it at $\sigma\to \infty$. 
Moreover, among the trajectories  there is 
a unique  one  $(\tilde
Z(\sigma), \tilde {\dot Z}(\sigma))$, with $\tilde{\dot Z}$  remaining finite at
$\sigma\to-\infty$.

In the original terms, this signifies  two important properties of the solutions
$X_0(t)$. First, all of them  ripple around  the exceptional solution
\begin{equation}
\bar X_0  = \ln \left( {3t} \right)
\end{equation}
approaching the latter at $t\to\infty$.  Second, the regular  at the
origin solution $\tilde X_0$ is unique and,  supplemented by condition
$\tilde X_0(0)=0$, should  look at $t\to 0$
as given by~(\ref{analyt}) with $\varepsilon_h=0$:
\begin{equation}\label{analyt0}
\tilde X_0  = t - \frac{3}{{10}}t^2  + \frac{4}{{35}}t^3 .
\end{equation}
This gives the qualitative picture of the looked-for regular solution $\tilde
X_0$ as a whole. 
The same concerns the regular $\tilde a_1=\tilde X_0$.
 As for  the regular $\tilde c_1$, it
have to approach asymptotically the exceptional solution 
\begin{equation}
\bar c_1  = 2 - \ln \left( {3t} \right)
\end{equation}
and  be approximated at $0\leq t<1$   by the power series 
\begin{equation}
\tilde c_1  =  - \frac{1}{{10}}t^2  + \frac{2}{{35}}t^3. 
\end{equation}

\section{Numerical analysis}

Here, we   numerically verify and visualize  the above analytical statements
of~\cite{Pir}.
The phase plane of  system~(\ref{1order}) is shown below:
\begin{center}
\includegraphics{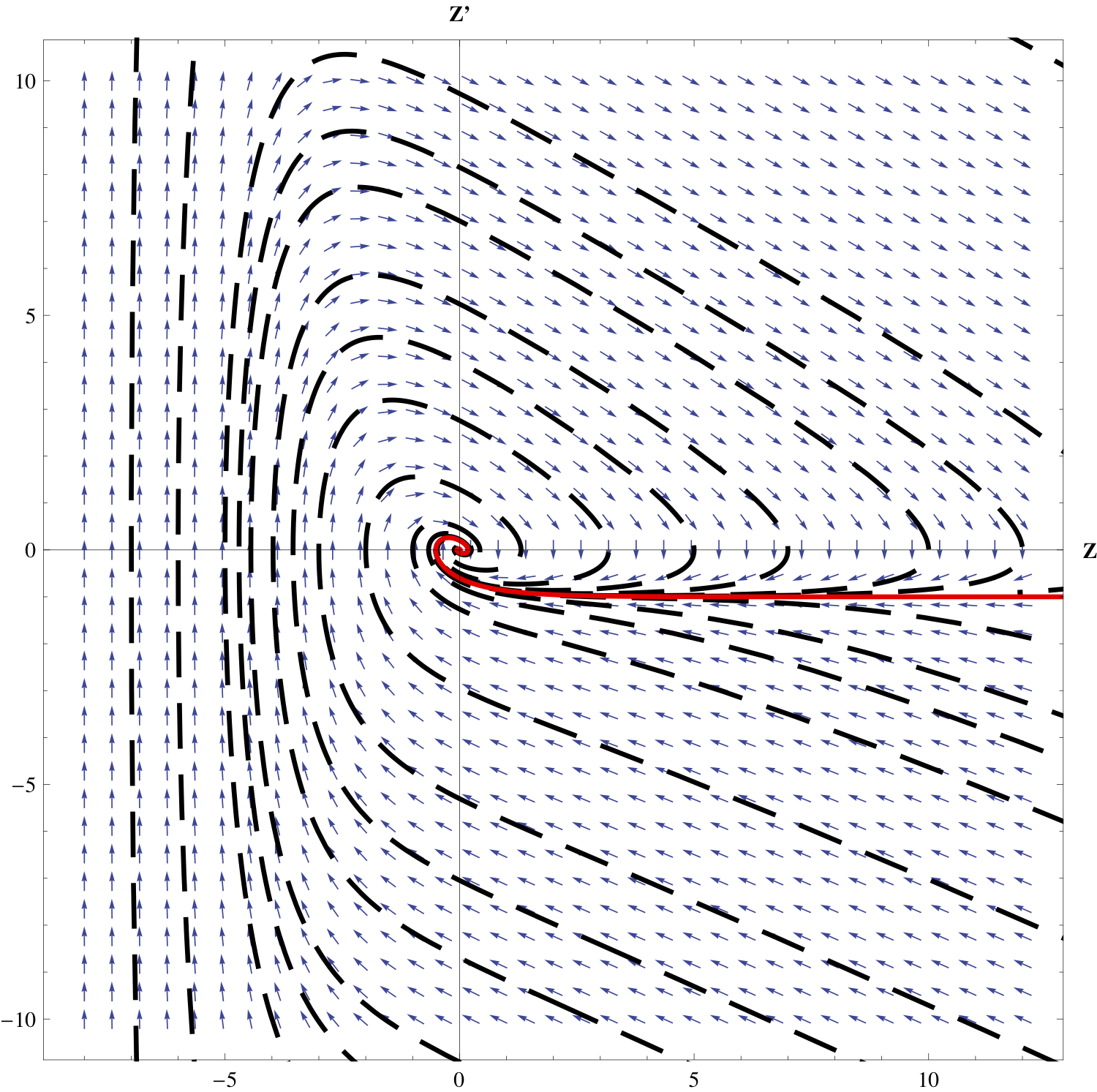}
\end{center}
The arrows present the direction field given by $d\dot Z/dZ   =  (d\dot
Z/d\sigma)/(dZ/d\sigma) $. The trajectories  are the lines $\dot Z(Z)$
tangential to the direction field in every point. 
The bulk of trajectories (dashed black lines) possesses $\dot Z\to -\infty$ at
$\sigma \to - \infty $  resulting in  the singular at $t \to 0$ solutions. 
There is just one  trajectory, with the finite $\dot Z$ 
($\dot Z\to -1$ at $\sigma \to - \infty$ ),  indicated by the solid red line. 
It may be associated with   solution ${\tilde X_0}$  regular at
$t=0$. The exceptional trajectory $(\bar Z=\bar{\dot Z}=0)$  corresponds to
exceptional solution $\bar X_0$. The picture above explicitly supports the 
statements  made in~\cite{Pir}.

To find the exact form of  regular  solution  we integrate the first
equation of  system~(\ref{MGE0}) numerically, with  boundary condition
taken from (\ref{analyt0}). This gives
\begin{center}
\includegraphics{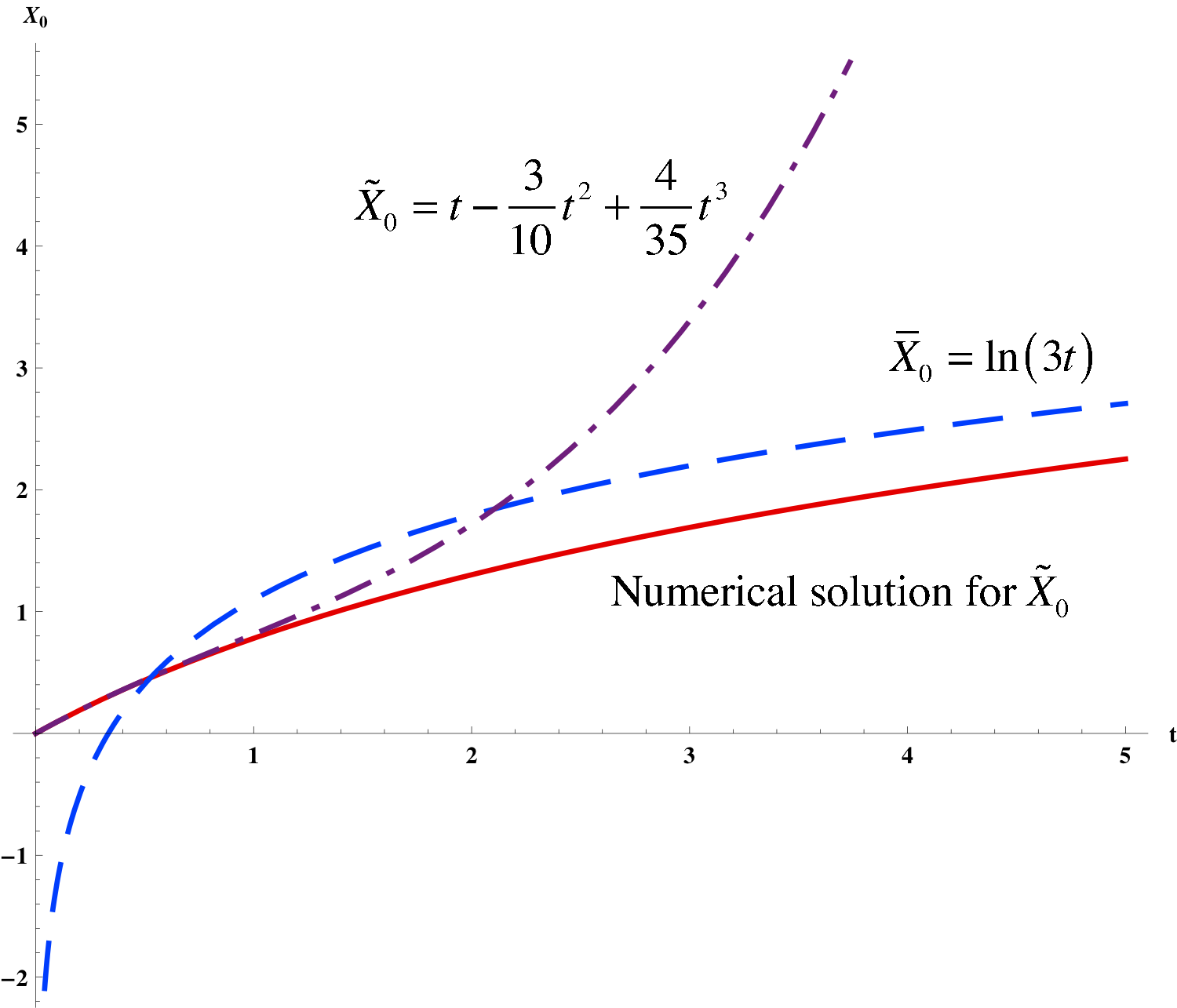}
\end{center}
It is well seen that the power series approximates  numerical
solution very good in  interval $0 \le t < 1$.
Asymptotically, the regular  solution $\tilde X_0$ vs.\  the exceptional  one
$\bar X_0$ looks~like:
\begin{center}
\includegraphics{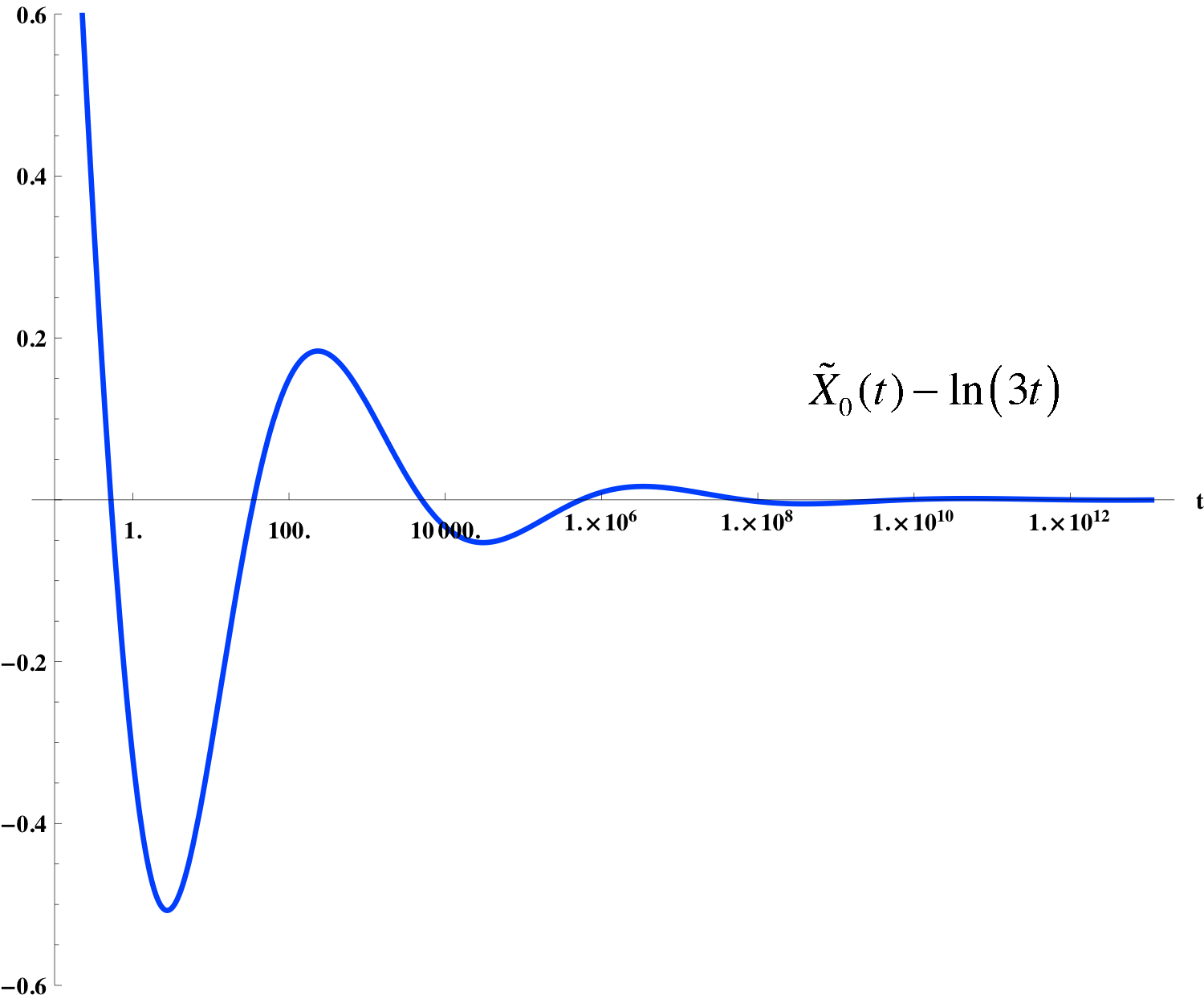}
\end{center}
Clearly, $\tilde X_0$ ripples around  $\bar X_0$ approaching the latter  
at $t\to\infty$.  The picture shows that this approach is though extremely slow.

The same concerns  $\tilde c_1$. Integrating the third equation of
system (\ref{MGE0}) numerically  we get  solution as follows:
\begin{center}
\includegraphics{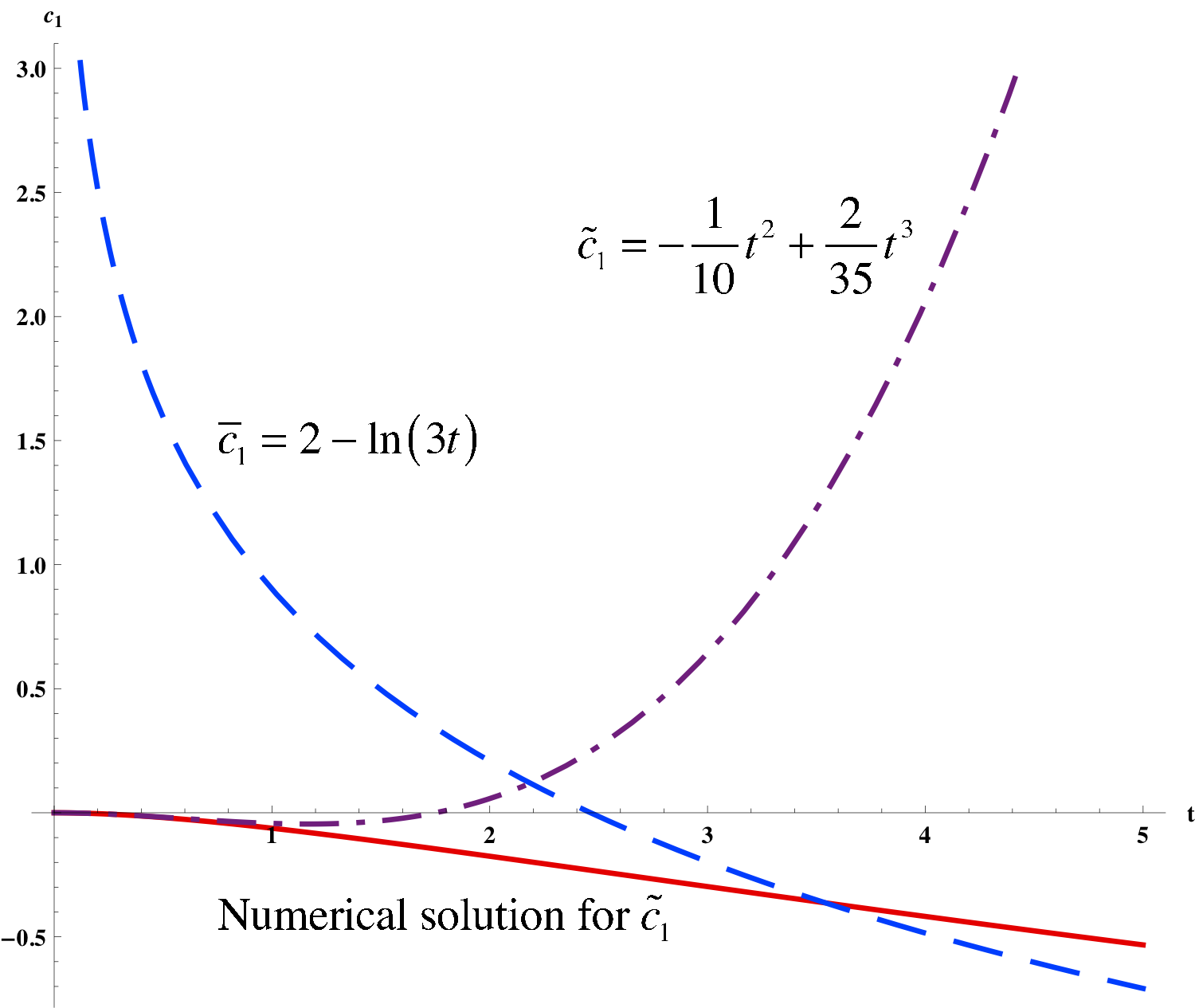}
\end{center}
At  large $t$, the solution also ripples  around the exceptional one  
approaching the latter asymptotically:
\begin{center}
\includegraphics{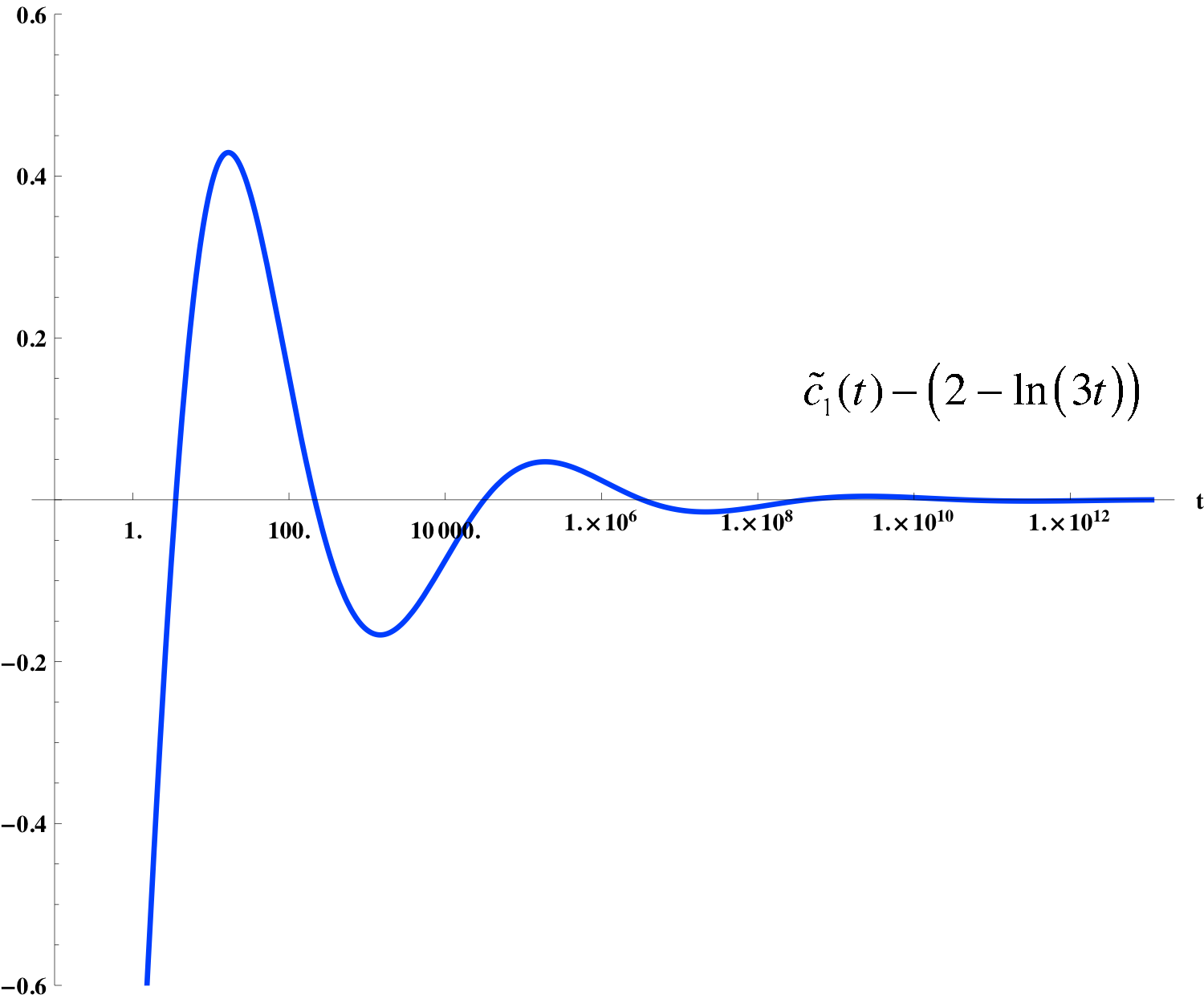}
\end{center}
Thus,  numerical study  totally confirms all the qualitative statements
about  regular solution made in \cite{Pir}.

At last,  to check the accuracy of  numerical calculations we input the found
numerical solutions into  consistency condition given by the last equation of
(\ref{MGE0}). The absolute (the difference of  L.H.S.\ and R.H.S.) and
relative (the ratio of  L.H.S.\ and R.H.S.\ minus unity) errors of
calculations are shown, respectively,   by the blue and red lines below:
\begin{center}
\includegraphics{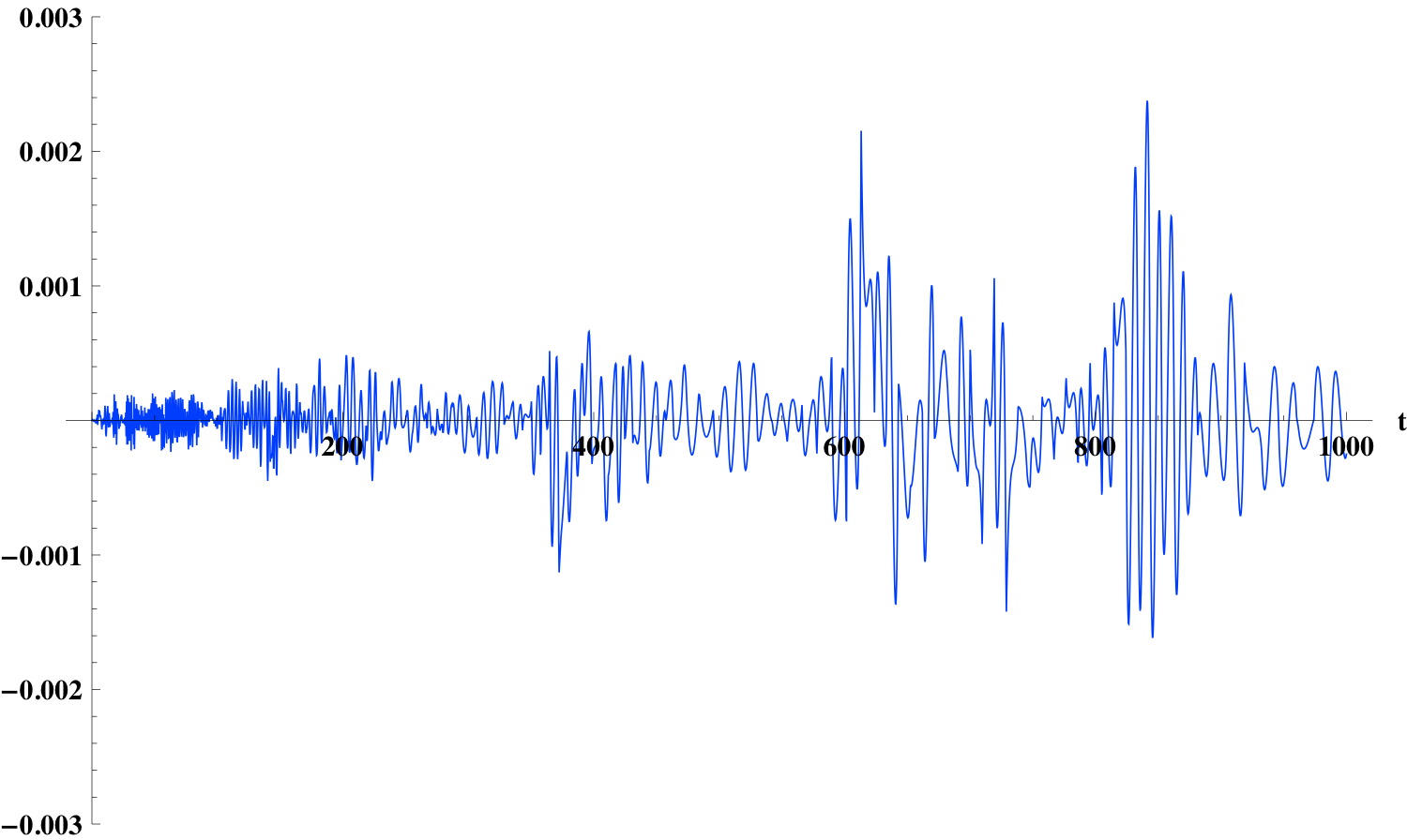}
\includegraphics{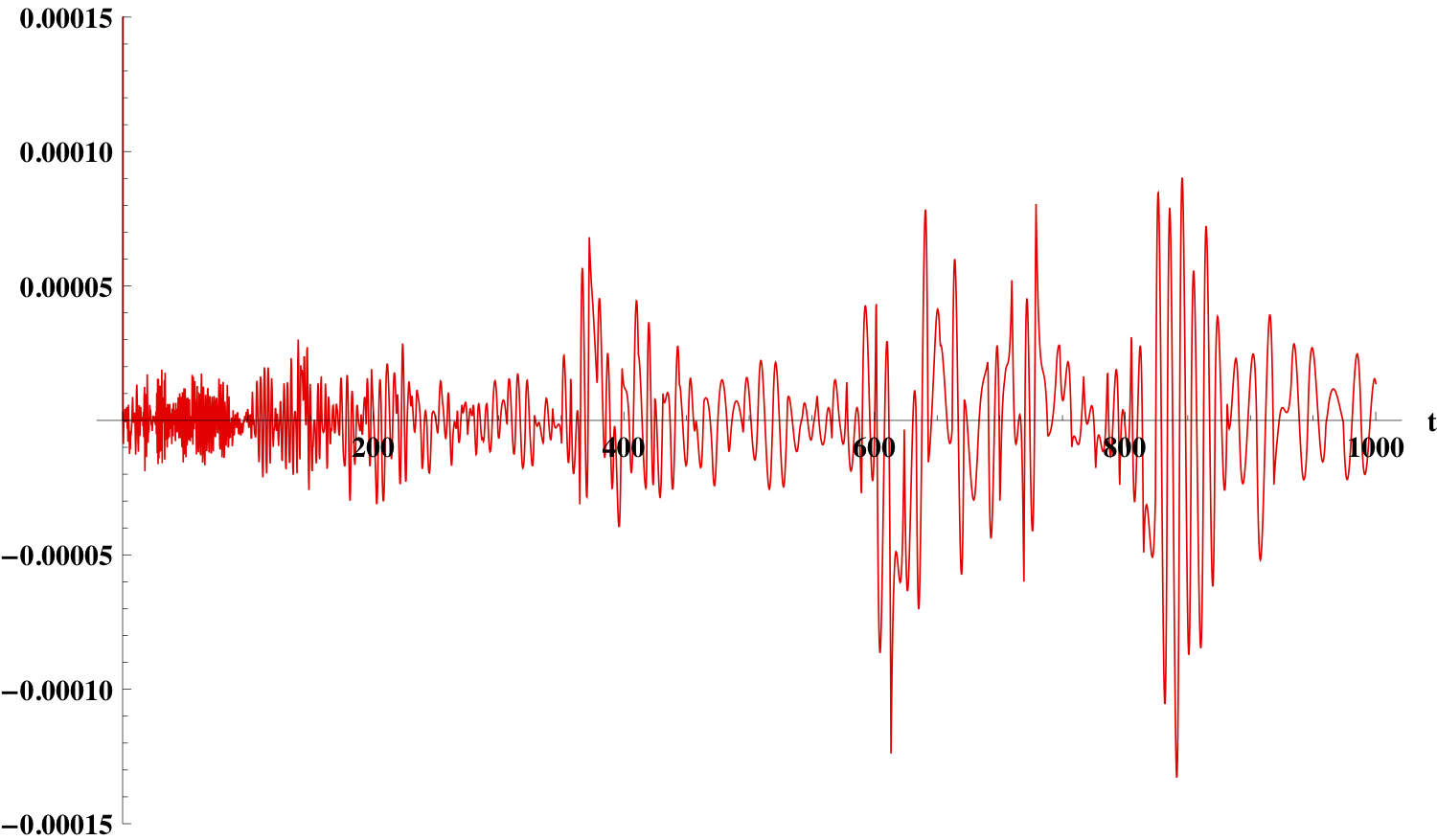}
\end{center}
Evidently, the achieved accuracy of numerical calculations is quite nice.

\section{Symbolic calculations}

Here, we examine the exact regular solution to  exact
system~(\ref{MGE}) in  a vicinity of $t=0$ under an arbitrary $\varepsilon_h$.
To this end, decompose the  looked-for exact solution $\tilde X(t)$ as  the
power series of $t$   as follows:                    
\begin{equation}
\tilde X = \sum\limits_{n = 0}^\infty  {t^n } \alpha _n(\varepsilon_h) ,
\end{equation}
with $\alpha_h$ being some  parameters ($\alpha_0=0$).
We proceed similarly   with $a(t)$  and  $c(t)$).
Implementing a system of symbolic calculations  and following the cyclic
perturbative procedure proposed in~\cite{Pir} we get   in a reasonable time
the 24 terms of  decomposition,  the first five of them being shown below:
\begin{eqnarray}
\alpha _1 & =& 1 , \nonumber\\
\alpha _2 & =&  - \frac{1}{{10}}\left( {3 + 5\varepsilon _h ^2 } \right),
\nonumber \\
\alpha _3 & =& \frac{1}{{105}}\left( {12 + 41\varepsilon _h ^2  
+ 35\varepsilon ^4 _h } \right), \nonumber \\
\alpha _4  &=&  - \frac{1}{{6300}}\left( {305 + 1573\varepsilon _h ^2  
+ {\rm{2735}}\varepsilon ^4 _h  + {\rm{1575}}\varepsilon ^6 _h } \right) ,
\nonumber\\
\alpha _5 & =& \frac{1}{{{\rm{173250}}}}\left( {{\rm{3774}} 
+ {\rm{25969}}\varepsilon _h ^2  + {\rm{68120}}\varepsilon ^4 _h  +
{\rm{79675}}\varepsilon ^6 _h  + {\rm{34650}}\varepsilon ^8 _h } \right)
\end{eqnarray}
(and similarly for $a_1$ and $c_1$).
The first three terms above reproduce  those given by~(\ref{analyt}).

Putting  now $\varepsilon _h = 0$ we can compare $\tilde X_0$
found previously with the present solution, two approximations to which being
shown below:
\begin{center}
\includegraphics{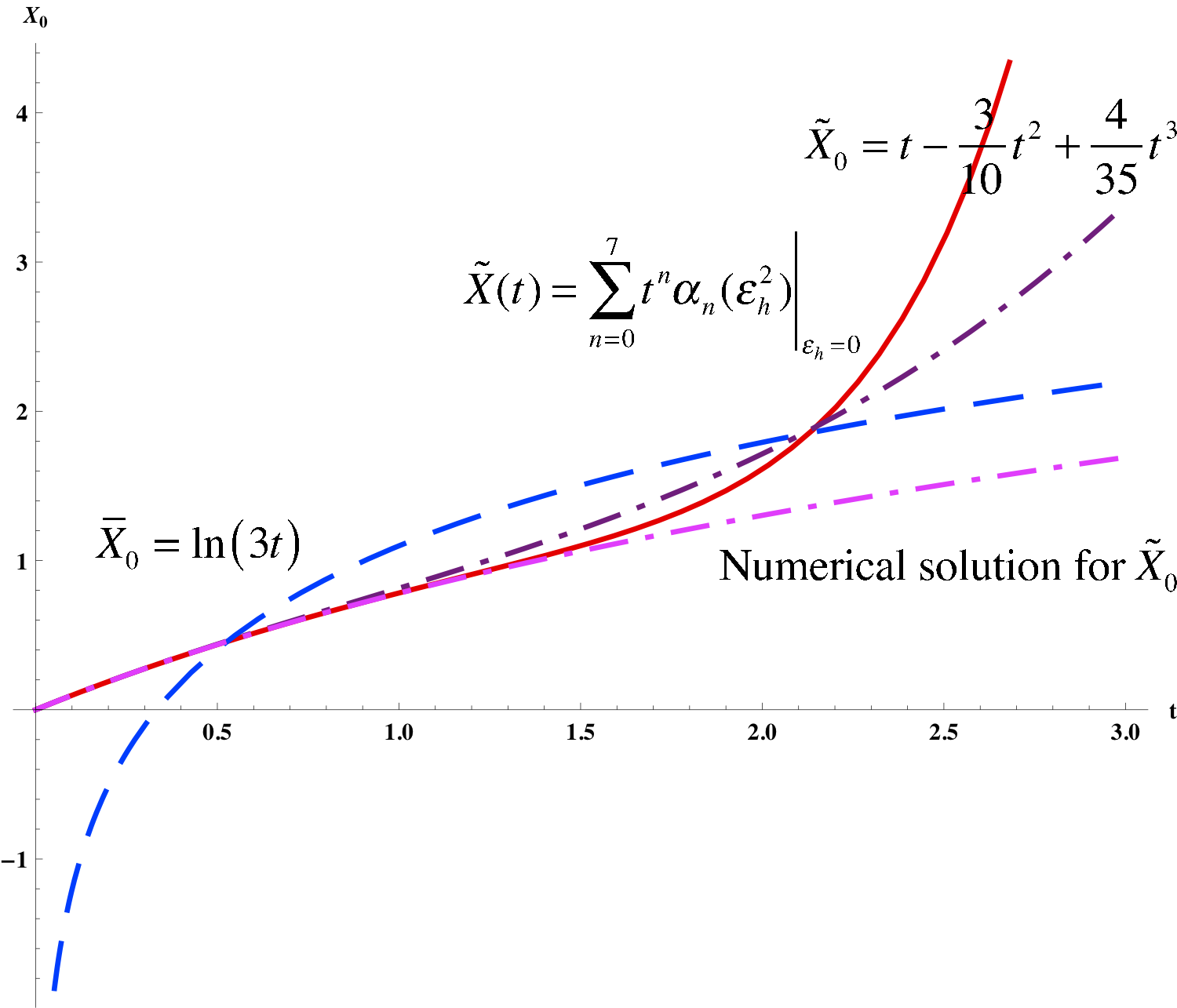}
\includegraphics{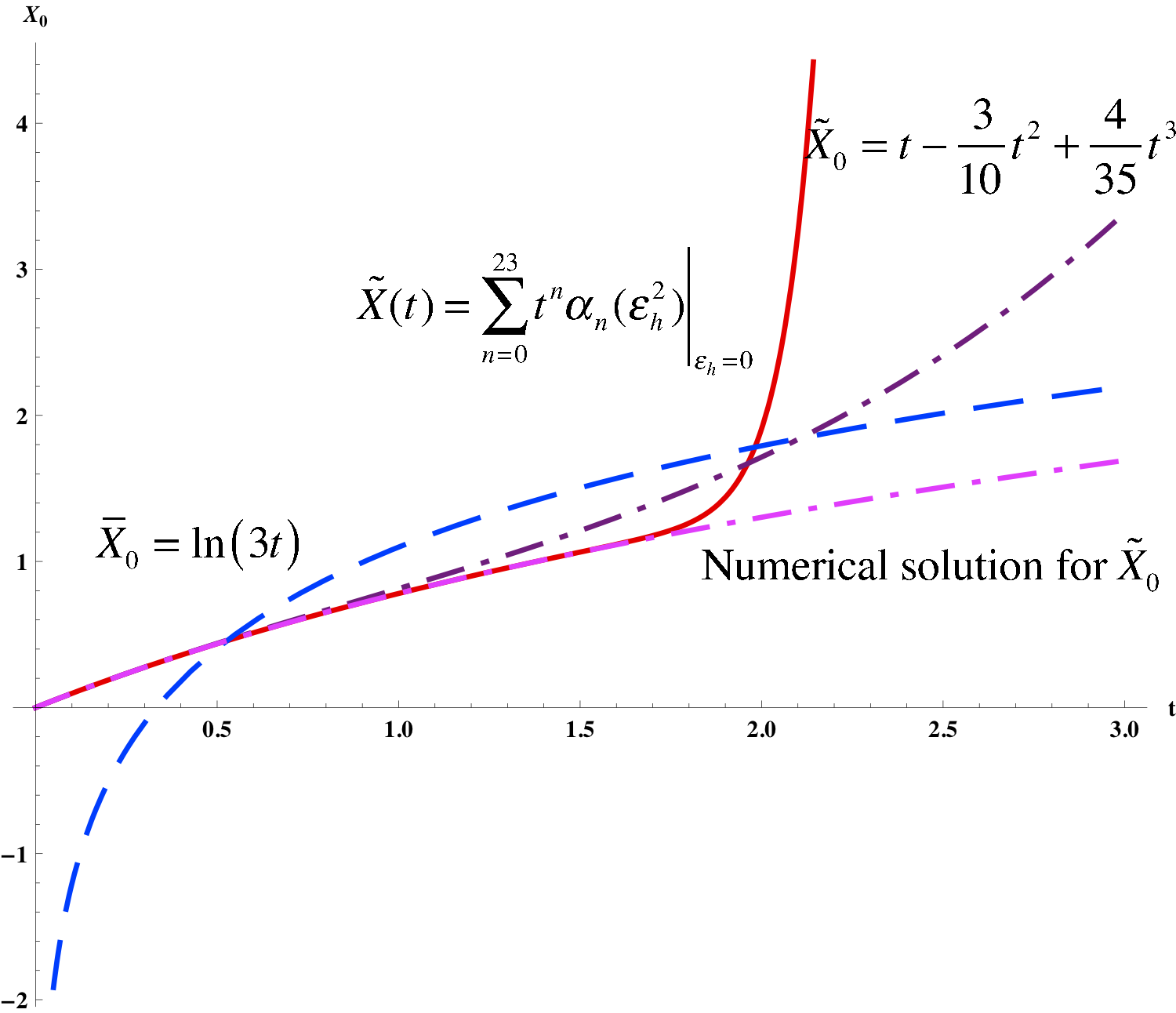}
\end{center}
It is  seen that the  power-series solution $\tilde X|_{\varepsilon_h=0}$
perfectly  matches the numerical one $\tilde X_0$ up to $t \leq 1.7$.  With $n$
increasing, the approximation clearly improves within this region but worsens
beyond it. The further analysis reveals the same picture with
$\varepsilon _h\neq 0$.  Fixing  a numerical value for $\varepsilon _h$ allows
one to calculate much more coefficients $\alpha _n$ in a  reasonable time. 
Thus, calculations with $\varepsilon_h=0$ up to 
$n \simeq 3 \cdot 10^7$ make it highly plausible that  the area of convergence 
of  power series is limited in this case to $t\leq 1.8$. It seems that
similar statement survives with an arbitrary  $\varepsilon_h<{\cal O}(1)$.

\section{Dark halos}

Here, we examine  validity of  analytical results for  rotation
velocity and the ensuing dark matter profile  found in~\cite{Pir}.
The circular rotation velocity of a test particle in the spherically symmetric
metric (\ref{ds2}) is, generally, as follows:
\begin{equation}
v^2 (r) = \frac{d}{{dr}} a(r)\Big/
 \frac{d}{dr} \ln \left( {r^2 c(r)}\right).
\end{equation}
With account for $a_1=X_0$,  the respective
velocity squared profile (in terms of $\xi=r/R_h$)  in the leading
$\varepsilon_h$-order looks like
\begin{equation}
v_h ^2 (\xi )/\varepsilon ^2 _h  \equiv U_h (\xi )= (\xi/2) dX_0 (\xi )/ d\xi .
\end{equation}
The regular solution  $\tilde X_0$ results then in: 
\begin{equation}
\tilde U_h (\xi )  =   \left\{
{\begin{array}{*{20}l}
{\xi ^2  - \frac{3}{5}\xi ^4  + \frac{{12}}{{35}}\xi ^6 ,} & 
{0 \le \xi  <1}, \\
{1,} & {\xi  \gg 1} , \\
\end{array}} \right.
\end{equation}
This analytical approximation vs.\  numerical result   is shown below:
\begin{center}
\includegraphics{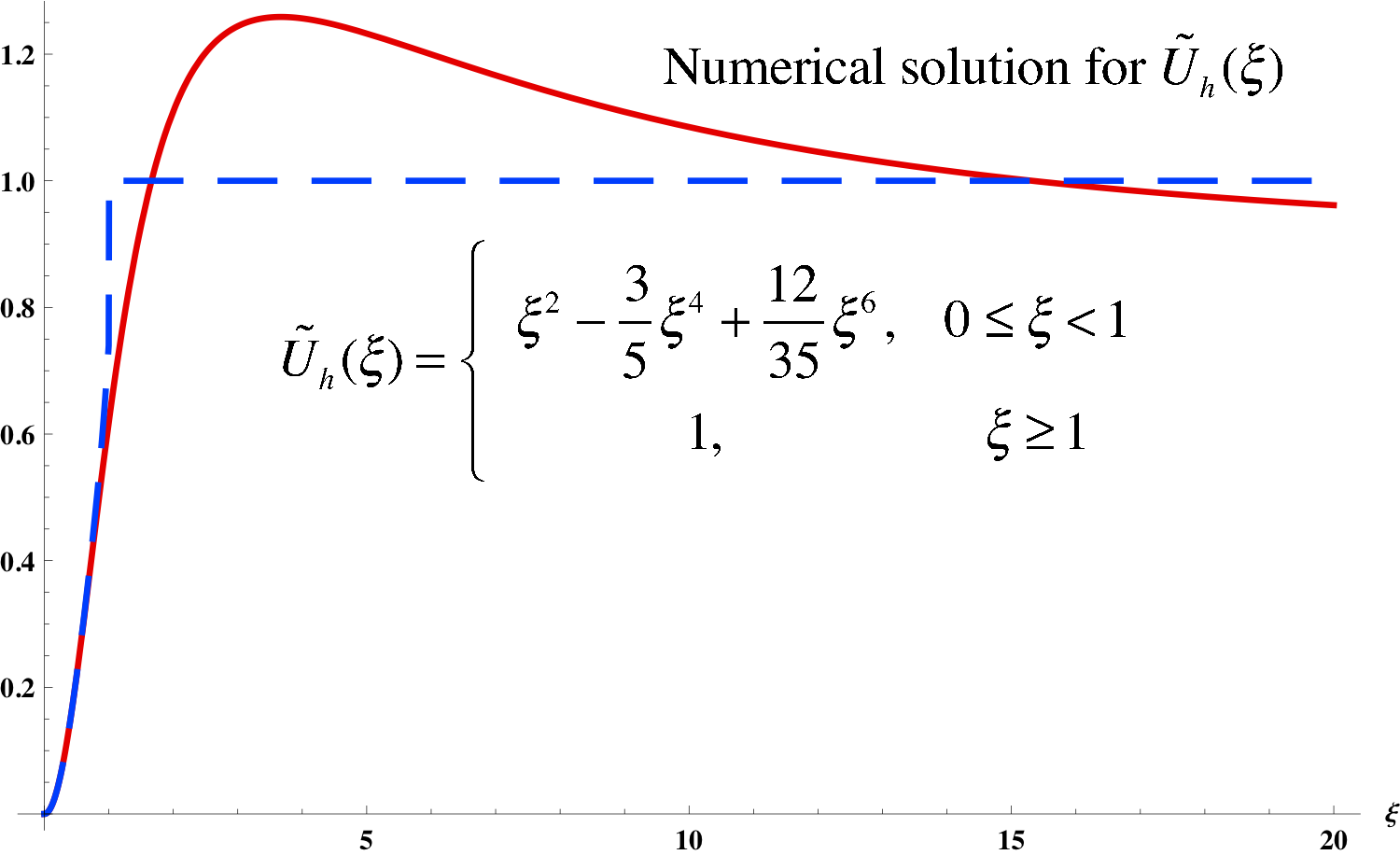}
\end{center}
with   the numerical $\tilde U_h$   approaching unity  asymptotically like 
\begin{center}
\includegraphics{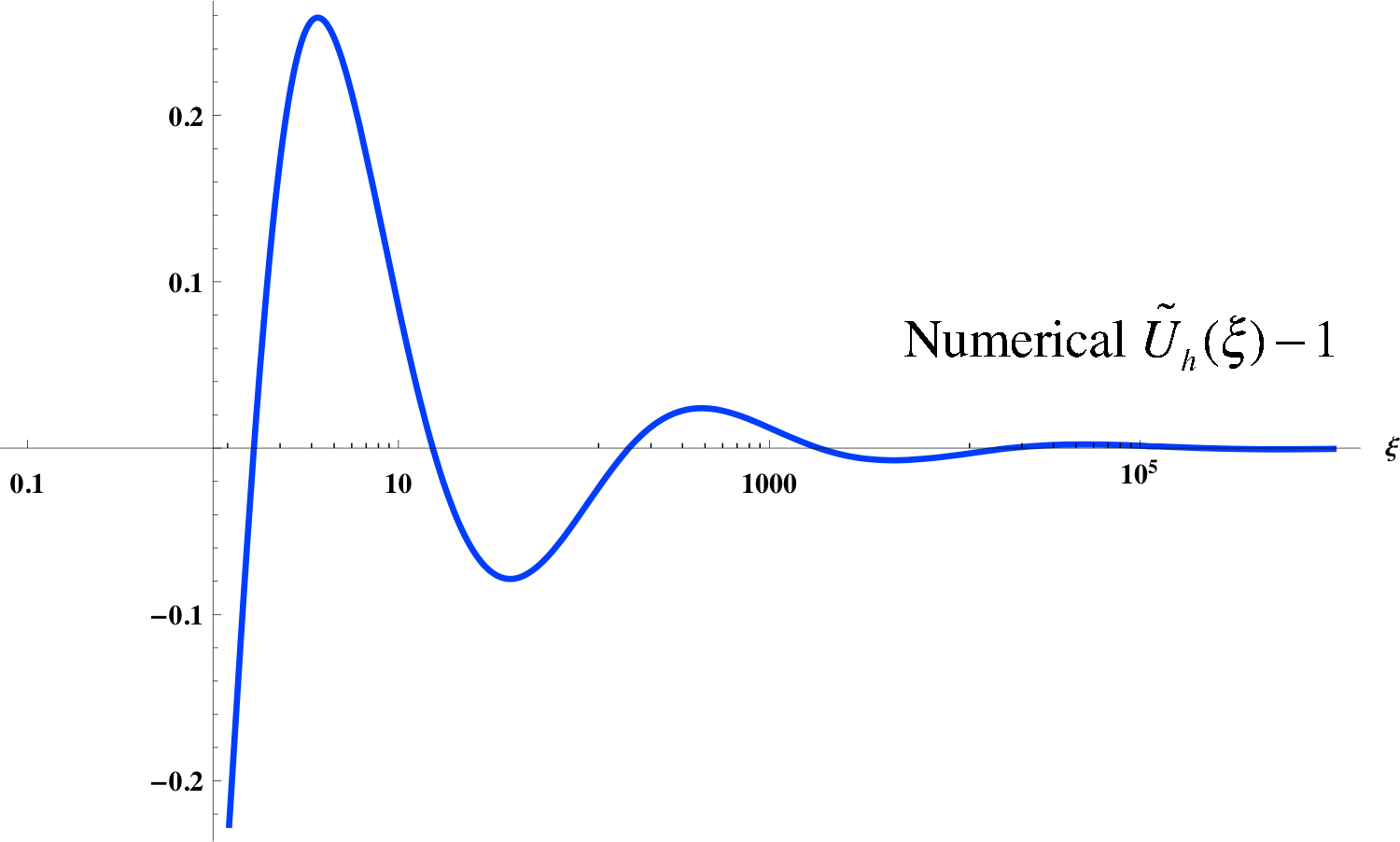}
\end{center}

The regular solution implies  the dark matter profile as follows: 
\begin{equation}
\rho _h (\xi ) / \rho _h (0)\equiv  \tilde P_h (\xi ) = \exp \left( { -
\tilde X_0  (\xi )} \right),
\end{equation}
where $\rho _h (0) =
6\varepsilon ^2 _h \kappa^2 _g /R^2 _h$.
Analytically, this results in  the soft-core halo  profile~\cite{Pir}:
\begin{equation}
\tilde P_h (\xi ) = \left\{ {\begin{array}{*{20}l}
  {1 - \xi ^2  + \frac{4}{5}\xi ^4 ,} & {0 \le \xi  < 1},  \\
 {\frac{1}{3}\xi ^{ - 2} ,} & {\xi  \gg 1},  \\
\end{array}} \right.
\end{equation}
with the finite central density $\rho _h (0)$. 
The analytical approximation vs.\  numerical result is as follows:
\begin{center}
\includegraphics{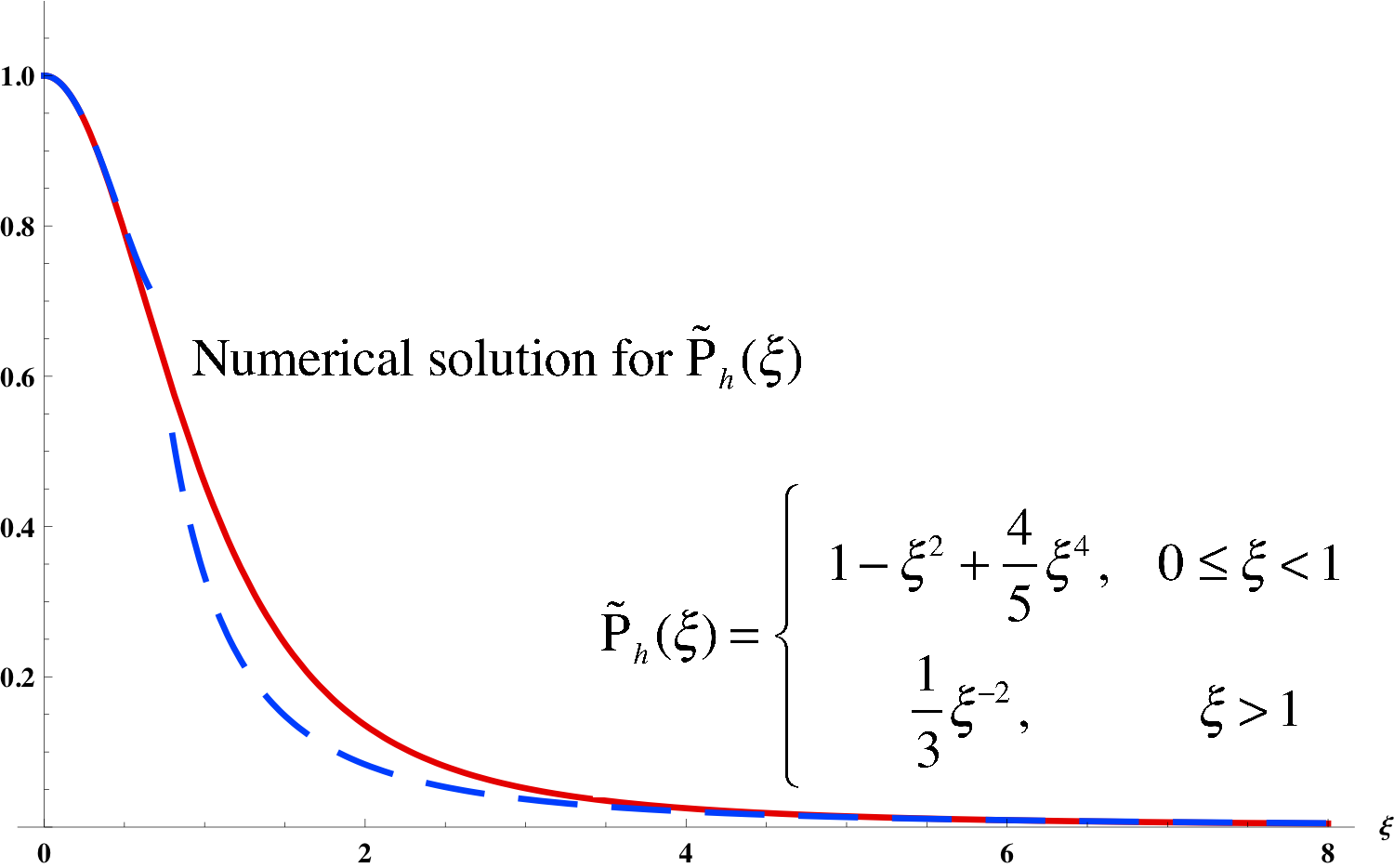}
\end{center}
Clearly, the accuracy of  analytical study is quite reasonable.

\section{Conclusion}

Summarizing,   numerical analysis of  regular solution to the
static spherically symmetric  equations of the  unimodular extended gravity in
empty space totally confirms and somewhat refines analytical statements
made in~\cite{Pir}. As a by-product,  symbolic calculations make it
highly plausible that the power-series decomposition of  solution is valid
just interior to a finite convergence radius. We are going to expand the 
conducted study on  the general solutions to the aforementioned equations, with
a view to refine the application  of  theory  to the galaxy  dark halos
started in~\cite{Pir}.

\end{document}